\documentclass[aps,11pt,nofootinbib,showpacs,preprintnumbers,prd,superscriptaddress]{revtex4}
\usepackage{epsfig,amssymb,amsmath,latexsym}
\newcommand\ep{\epsilon}

\newcommand\beq{\begin{equation}}
\newcommand\eeq{\end{equation}}
\newcommand\bea{\begin{eqnarray}}
\newcommand\eea{\end{eqnarray}}
\newcommand\bi{\begin{itemize}}
\newcommand\ei{\end{itemize}}
\newcommand\ben{\begin{enumerate}}
\newcommand\een{\end{enumerate}}

\newcommand\etal{{\hbox{{\textit\ et. al.\/}\textit\ }}}

\def\dfrac#1#2{{\displaystyle\frac{#1}{#2}}}

\newif\ifboo \boofalse

\def\lsim{\mathrel{\rlap{\lower4pt\hbox{\hskip1pt$\sim$}}
    \raise1pt\hbox{$<$}}}         %less than or approx. symbol
\def\gsim{\mathrel{\rlap{\lower4pt\hbox{\hskip1pt$\sim$}}
    \raise1pt\hbox{$>$}}}         %greater than or approx. symbol
\preprint{RECAPP-HRI-2009-016}
\begin{document}
 \textheight=23.8cm
\title{\Large{Maximal mixing  as a `sum' of small mixings}}
\date{\today}
\author{\bf Joydeep Chakrabortty}
 \email{joydeep@hri.res.in}
\affiliation{Harish-Chandra Research Institute, Allahabad 211 019,
India}
\author{\bf Anjan S. Joshipura}
\email{anjan@prl.res.in} \affiliation{Theory Group, Physical
Research Laboratory, Ahmedabad 380 009, India}
\author{\bf Poonam Mehta}
 \email{poonam@rri.res.in}
\affiliation{Theoretical Physics Group, Raman Research Institute,
Bangalore 560 080, India}
\author{\bf Sudhir K. Vempati}
\email{vempati@cts.iisc.ernet.in} \affiliation{Centre for High
Energy Physics, Indian Institute of Science, Bangalore 560 012,
India}
\begin{abstract}
In models with two sources of neutrino masses, we look at the
possibility of generating maximal/large mixing angles in the total
mass matrix, where both the sources have only small mixing angles.
We show that in the two generation case, maximal mixing can
naturally arise only when the total neutrino mass matrix  has a
quasi-degenerate pattern.  The best way to demonstrate this is by
decomposing the quasi-degenerate spectrum in to hierarchial and
inverse-hierarchial mass matrices, both with small mixing. Such a
decomposition of the quasi-degenerate spectra  is in fact very
general and can be done irrespective of the mixing present in the
mass matrices. With three generations, and two sources, we show that
only one or all the three small mixing angles in the total neutrino
mass matrix  can be converted to
 maximal/large mixing angles.  The decomposition of the degenerate pattern
 in this case is best realised in to sub-matrices whose dominant eigenvalues
 have an alternating pattern.  On the other hand,  it is possible to generate
 two large and one small mixing angle if either one or both of the
sub-matrices contain maximal mixing. We present example textures of
this. With three sources of neutrino masses, the results remain
almost the same as long as all the sub-matrices contribute equally.
The Left-Right Symmetric model where Type I and Type II seesaw
mechanisms are related provides a framework where small mixings can
be converted to large mixing angles, for degenerate neutrinos.
\end{abstract}

\maketitle
\vskip .6 true cm
%------------------------------------------------------------------------
\section{Introduction}
 \label{intro}
 While  neutrino masses have been thoroughly established
experimentally~\cite{conchamaltoni}, the question of how they attain
their masses still needs to be understood.  Perhaps, the most
elegant mechanism of generating neutrino masses is through the
seesaw mechanism~\cite{seesaw}.  Here
 one trades the tininess of the neutrino masses with high scale
 Majorana masses  for right-handed neutrinos introduced for this purpose.
While the original seesaw mechanism dealt only with right-handed
heavy neutrino states, in recent years, it has been realized that
there could be other heavy triplet scalars~\cite{seesaw2} or even
triplet fermions~\cite{seesaw3,goran} which could play the same role
as right-handed neutrinos in the original seesaw mechanism. These
mechnaisms are named as Type I, Type II and Type III seesaw
mechanisms respectively (for recent reviews please see
~\cite{strumiareview,nirdavidsonreview}). While one of the three
seesaw mechanisms suffices to generate non-zero neutrino masses, it
is interesting to note that  in most Grand Unified Theory (GUT)
models, there is more than one seesaw mechanism  at work. For
example, in SO(10) models both Type I and Type II seesaw mechanisms
are simultaneously present as soon as one considers representations
of the type $\overline{126}$~\cite{babumohapatra}. In Left-Right
Symmetric (LRS) models, the Type I and Type II seesaw  mechanisms
are not just  present, but they are also related to each
other~\cite{akhmedov}. Similarly, Type I and Type III mechanisms
co-exist in SU(5) model with an adjoint fermion
representation~\cite{goran}. In most of these investigations,
 typically one considers one of them to be dominant while the other to be subdominant.

 One of the crucial features of seesaw mechanism was its ability to generate large
 or maximal mixing even though the mixing present in the Dirac neutrino Yukawa
 couplings is small like in the hadronic sector. In fact this is what typically
 happens in a SO(10) GUT~\cite{smirnovaf}, where neutrino Dirac Yukawa couplings
 have the same structure as the top Yukawa couplings; even in such cases large mixing
 in the neutrino sector is possible.  However this would
 require large hierarchies in the masses of the right-handed neutrinos which is
  in conflict with thermal leptogenesis in these models~\cite{thermallepto}~\footnote{This is
  true when the mixing angles in neutrino Dirac Yukawa are exactly like CKM angles.}.
 In the present work, we look for an alternative method to generate large/maximal
 mixings instead of using the `seesaw-effect'.  We will use the fact that most models
 GUT models like SO(10) have more than one seesaw mechanism at work.
 However, instead of restricting ourselves to any particular GUT model or the seesaw mechanism,
 we analyze the general situation where there are two sources for neutrino masses and both of
 these contain small neutrino mixing.

 Our analysis shows that the total neutrino mass matrix which is given by the
 sum of the two neutrino sources can have large or maximal mixing only if the
 resulting pattern of the neutrino masses is of the quasi-degenerate form.
 A crucial condition which needs to be satisfied to reach this conclusion is
 that the large eigenvalues of the sub-matrices do not cancel in the total
 mass matrix.  This results in the sub-matrices taking the form of hierarchial
 and inverse-hierarchial matrices whose sum leads to the quasi-degenerate
 form.  Given that the decomposition of the degenerate spectrum in to hierarchial
 and inverse-hierarchial mass matrices is quite generic,  as we will demonstrate
 here, one can enumerate the possible forms the individual sub-matrices can take.
It should be noted that the decomposition itself is independent of
the actual mechanism responsible for generating neutrino masses
\textit{i.e}, doesn't depend on whether there is a seesaw mechanism
at work or not.  It is well known that the  quasi-degenerate
 pattern  for neutrino masses can be achieved both with~\cite{joshipura} and
without~\cite{marajasekaran} seesaw mechanism.  However,  as will
demonstrate later, the model dependence enters, if one wants to
realise the decomposition in terms of independent Lagrangian
parameters which for example  is possible in Type I seesaw
mechanism.

The simple example where our scheme of things can be realised is the
LRS model where both Type I and Type II seesaw mechanisms are
simultaneously present. We will explicitly present the conditions on
the LRS parameters required in order to realize the mechanism. The paper is
organised as follows. In Sec.~\ref{sectwo}, we analyse the two
generation case and show how only when the quasi-degeneracy is
satisfied in the final matrix, one can have large or maximal mixing.
We also describe all the possible decompositions of the degenerate
spectra. We further discuss how this scheme can be incorporated
within the LRS models.  In Sec.~\ref{secthree}, we consider two
cases {\sl{(a)}} with two seesaw mechanisms or two sources, and
{\sl{(b)}} with three sources. We then demonstrate the decomposition
of the quasi-degenerate spectrum and discuss the subtleties which
arise in this case. We also determine the required parameter values within
 the LRS model for both the cases. We close with summary
 and outlook in Sec.~\ref{secfour}. Generalisation of our result to the
  case of  $n$ sources of neutrino masses is given in Appendix~\ref{appendix}.

%-----------------------------
 \section{Large mixing as sum of small mixing angles}
 \label{sectwo}
Consider a  model for neutrino masses in which the total neutrino
mass matrix is given by
\begin{equation}
{\mathbb M}_\nu = {\mathbb M}_{\nu}^{(1)} + {\mathbb M}_{\nu}^{(2)},
\end{equation}
where $ {\mathbb M}_{\nu}^{(1)}$ and $ {\mathbb M}_{\nu}^{(2)}$  can
be thought of as two individual sources of neutrino mass. For
example, $ {\mathbb M}_{\nu}^{(1)}$ could have its origin in Type I
seesaw whereas $ {\mathbb M}_{\nu}^{(2)}$ could have its origin in
Type II seesaw mechanism
 in a model like SO(10) where both these mechanisms are simultaneously
 present~\footnote{In fact, in most models of neutrino masses, one of them,
 say  $ {\mathbb M}_{\nu}^{(1)}$ could correspond to zeroth order mass
 while the other  $ {\mathbb M}_{\nu}^{(2)}$ could correspond to perturbations
 required to make contact with the experimental results.}. Irrespective of their
 origin, let us assume that both $ {\mathbb M}_{\nu}^{(1)}$  and
 $ {\mathbb M}_{\nu}^{(2)}$ contain only small mixing angles. We now
 ask the question whether it is possible to have in the total mass matrix
 $ {\mathbb M}_\nu$ {\sl{(a)}} maximal or large mixing, and {\sl{(b)}} a \textit{reasonable} $\Delta \mbox{m}^2$
 without fine-tuning. By this we mean, that the $\Delta m^2$ is determined in terms of the dominant
 eigenvalues of $ {\mathbb M}_{\nu}^{(i)}$ (where $i=1,2$). To make  the
  discussion concrete, we will stick to two generation case in
  the present section. Denoting
 \begin{equation}
 {\mathbb M}_{\nu}^{(1)} = \left( \begin{array}{cc}
m_{ee}^{(1)} & m_{e\mu}^{(1)} \\
  m_{e\mu}^{(1)} & m_{\mu\mu}^{(1)} \end{array} \right)  ,\;\;\;\;\;\;
  {\mathbb M}_{\nu}^{(2)} = \left( \begin{array}{cc}
m_{ee}^{(2)} & m_{e\mu}^{(2)} \\
  m_{e\mu}^{(2)} & m_{\mu\mu}^{(2)} \end{array} \right),
 \end{equation}
we can easily derive the following relations :
\begin{eqnarray}
\tan 2 \theta &=& {2  m_{e\mu}^{(1)}  + 2 m_{e\mu}^{(2)} \over  m_{\mu\mu}^{(2)} + m_{\mu\mu}^{(1)} -    m_{ee}^{(2)} - m_{ee}^{(1)} } \\
&=& \tan 2\theta^{(1)} {1 \over (1 + d)} + \tan 2 \theta^{(2)} {d
\over (1 + d)}~,
 \end{eqnarray}
where $ d = (m_{\mu\mu}^{(2)} - m_{ee}^{(2)})/( m_{\mu\mu}^{(1)} -
m_{ee}^{(1)} ) $ and $\theta^{(1)}$ and $\theta^{(2)}$ are the
mixing angles of ${\mathbb M}_{\nu}^{(1)}$ and  ${\mathbb
M}_{\nu}^{(2)}$ respectively. From this expression, it is obvious
that when both the mixing angles, $\theta^{(1)} $ and $\theta^{(2)}$
are small, the only region where $\theta$ would be maximal is when
$d = -1 $. Notice that the small mixing in ${\mathbb M}_{\nu}^{(i)}$
would mean {\sl{(a)}} 2 $m_{e\mu}^{(i)} \ll |m_{\mu\mu}^{(i)} -
m^{(i)}_{ee} |$, and {\sl{(b)}} $m^{(i)}_{\mu\mu} \neq m^{(i)}_{ee}$
for $i = (1,2)$ \textit{i.e,} the splitting in the diagonal entries
is much larger than the off-diagonal entry such that the mixing
remains small. Assuming at least one of the diagonal entries in each
of the matrix ${\mathbb M}_{\nu}^{(i)}$ is large, we have the
following three solutions for $d = - 1$  \begin{enumerate}
\item[] (A) $m^{(2)}_{\mu\mu} = -m^{(1)}_{\mu\mu}$~,
\item[] (B) $m^{(2)}_{ee} =
-m^{(1)}_{ee}$~, and
\item[] (C) $m^{(2)}_{ee} = m^{(1)}_{\mu\mu}$ or
$m^{(2)}_{\mu\mu} = m^{(1)}_{ee}$~.
\end{enumerate}
The solution of the type (A) would represent the case in which both
the matrices ${\mathbb M}_{\nu}^{(i)}$ are of the hierarchial form
with one dominant diagonal element (the $\mu\mu$ entry). However in
the total mass matrix ${\mathbb M}_{\nu}$ this entry gets cancelled.
To illustrate this, consider the following textures for ${\mathbb
M}_{\nu}^{(i)}$
\begin{equation}
{\mathbb M}_{\nu} = m_1 \left( \begin{array}{cc}
z & x \\
x & 1 + z' \end{array} \right)  + m_2 \left( \begin{array}{cc}
0 & y \\
y & -1 \end{array} \right)~,
\end{equation} where $x,y,z$ are the small entries compared to
$m_{\mu\mu}^{(1)}/m_1 \equiv (1 + z')$ and  $m_{\mu\mu}^{(2)}/m_2
\equiv -1 $. Notice that the dominant eigenvalues of  ${\mathbb
M}_{\nu}^{(i)}$ have opposite CP parities as the maximal mixing
requirement condition is now given as $m_1 \approx m_2 \approx m$.
In this limit, the total mass matrix has the form
\begin{equation}
{\mathbb M}_{\nu} = m \left( \begin{array}{cc}
z & x + y \\
x+ y & z' \end{array} \right)~.
\end{equation}
 The total mass matrix here has no trace of the dominant element of the $\mathcal{O}(m)$ which was present in the sub-matrices.
 It has been cancelled in such a way that the condition $m^{(2)}_{\mu\mu} +  m^{(1)}_{\mu\mu}  =  m^{(1)}_{ee} + m^{(2)}_{ee}$ is satisfied,
 which is the same as the full condition of $d = -1$ which would mean $z' = z$,
 rather than the sub-condition, (A) $m^{(2)}_{\mu\mu} = -m^{(1)}_{\mu\mu}$, which would instead mean $z' = 0$.
 The role of the large element  of the  $\mathcal{O}(m)$ has only been to generate the small mixing
 in the respective  ${\mathbb M}_{\nu}^{(i)}$.  Thus, at the level of total mass matrix ${\mathbb M}_{\nu}$, the properties
 are determined by the small entries of the original sub-matrices.
 The mass-squared splitting $\Delta \mbox{m}^2$ of the total mass matrix in terms of the elements of ${\mathbb M}_{\nu}^{(i)}$ is
 given by
 \begin{equation}
\Delta \mbox{m}^2 = ( m^{(1)}_{ee} + m^{(1)}_{\mu\mu} + m^{(2)}_{ee}
+ m^{(2)}_{\mu\mu})  \sqrt{ ( m^{(1)}_{\mu\mu} + m^{(2)}_{\mu\mu} -
m^{(1)}_{ee} - m^{(2)}_{ee})^2 + 4 (m^{(1)}_{e\mu} + m^{(2)}_{e\mu}
)^2}~,
\end{equation}
which reduces in the present case to
\begin{equation}
\Delta \mbox{m}^2 = m^2 (z + z') \sqrt{ 4 (x + y)^2 + (z -z')^2}~.
\end{equation}
In the limit $z'  \rightarrow 0$, the mixing $\tan 2 \theta = 2 (x +
y)/z $ would depend on the relative magnitudes of $x,y$ and $z$ with
large mixing being possible as long as $x + y \gg z$. Similarly, in
the limit $z' \approx z$, maximal/large mixing is possible, and a
hierarchial pattern for the neutrinos can arise if $x, y \sim z,
z'$.  Solutions of the class (B) also lead to similar results with
the dominant entries of the sub-matrices ${\mathbb M}_{\nu}^{(i)}$
being cancelled in the total mass matrix.  We do not find these
solutions attractive as large mixing can only come when the dominant
elements (`$ee$' elements in this case) cancel precisely to such an
extent to be equal to the sum of the other diagonal elements
(`$\mu\mu$' elements). We now go on to discuss the solutions (C)
which we find  more natural.

The solutions of the type (C) are given by $m^{(2)}_{ee} =
m^{(1)}_{\mu\mu}$ or  $m^{(2)}_{\mu\mu} = m^{(1)}_{ee}$.  The
condition now requires that the opposite diagonal elements of the
sub-matrices are equal. This naturally sets the $ {\mathbb
M}_{\nu}^{(i)}$ to have an opposite ordering of their eigenvalues
\textit{i.e,} one with normal hierarchy and the other has inverse
hierarchy.  For illustration, let us consider the  (sub)-case with
$m^{(2)}_{\mu\mu} = m^{(1)}_{ee}$. This can be represented as
\begin{eqnarray}
{\mathbb M}_{\nu} = {\mathbb M}^{\rm{(1)}}_{\nu} + {\mathbb M}
^{\rm{(2)}} _{\nu} &=&
 m_1  \begin{pmatrix}
0 & x \\
x & 1
 \end{pmatrix}  + m_2 \begin{pmatrix}
1  & - y \\
-y &  0
 \end{pmatrix}
 \nonumber\\
&=& \begin{pmatrix}
 m_2   & m_1 x - m_2 y  \\
m_1 x - m_2 y  & m_1
 \end{pmatrix}~,
 \label{massflsmallsum}
\end{eqnarray}
where $x,y$ are small entries with $m_{\mu\mu}^{(1)} \equiv m_1$ and
$m_{ee}^{(2)} \equiv m_2$. Note that here too as in the earlier case
the mixing angles in the individual sub-matrices are small, $\theta
\simeq$ $x$ or $y$, where as the total mixing matrix is given by
\begin{equation}
\tan 2 \theta = {2 (m_1 x - m_2 y) \over m_1 - m_2}~.
\end{equation}
In the limit of exact degeneracy between $m_1$ and $m_2$, the mixing
is maximal as is evident. However, an important assumption is that
both the $m_1$ and $m_2$ carry the same sign or equivalently have
the same CP parity~\footnote{If the CP parities are opposite the
mixing will remain small.}.  In a more general situation, say when
the zeros of the matrices on the RHS are of
Eq.~(\ref{massflsmallsum}) are filled with small entries (`$ee$'
element in $ {\mathbb M}_{\nu}^{(1)}$ and `$\mu\mu$' element in  $
{\mathbb M}_{\nu}^{(2)}$), the condition for the
 large mixing is given by   $|m_1 - m_2| < 2 (m_1 x - m_2 y)$.
 Thus, the splitting in the diagonal
 entries should be much smaller than the off-diagonal elements.
 The spectrum of the total mass
 matrix points towards a quasi-degenerate pattern. The eigenvalues are given by :
\begin{equation}
 \lambda_{1,2} = \dfrac{1}{2}
\left[m_1+ m_2  \mp \sqrt{(m_1 - m_2 )^2 + 4 (m_1 x - m_2 y)^2} \right],
\end{equation}
which in the limit $m_1 \approx m_2 \approx m$ take the form $m-
\epsilon, m + \epsilon$, with $\epsilon  =  m (x-y) $ being the
order of the off-diagonal entry. The $\Delta \mbox{m}^2 = 4 m
\epsilon$. The other solution of (C), $m^{(2)}_{ee} =
m^{(1)}_{\mu\mu}$, corresponds to an interchange of $m_1$ and $m_2$
and would lead to similar conclusions.

Finally, let us consider a class of solutions with two large
diagonal entries in each of the $ {\mathbb M}_{\nu}^{(i)}$. However
given that the mixing in each of them is small, as per the
discussion above, the splitting between the diagonal elements should
be larger than the off-diagonal entry. This can be parameterised by
the following set of matrices :
\begin{eqnarray}
{\mathbb M}_{\nu} = {\mathbb M}^{\rm{(1)}}_{\nu} + {\mathbb M}
^{\rm{(2)}} _{\nu} &=&
 m_1  \begin{pmatrix}
1 + \rho & x \\
x & 1
 \end{pmatrix}  + m_2 \begin{pmatrix}
1  & x' \\
x' &  1+ \rho'
 \end{pmatrix}
 \nonumber\\
&=& \begin{pmatrix}
 m_1( 1 + \rho) +  m_2   & m_1 x  + m_2 x'  \\
m_1 x' + m_2 y  & m_1 + m_2 ( 1 + \rho')
 \end{pmatrix},
 \label{2dsumdeg}
\end{eqnarray}
where $x,x'$ are small entries compared to one as before and $\rho,
\rho'$ are chosen such that $2 |x/\rho| \ll 1$ and $2 |x' /\rho'|
\ll 1 $ to keep the mixing small in $ {\mathbb M}_{\rm{i}}^\nu$.
This would  mean a relative hierarchy of the elements in the
individual matrices, $m_{ee}^{(1)} \gg m_{\mu\mu}^{(1)} \gg
m_{e\mu}^{(1)}$ in $ {\mathbb M}_{\rm{1}}^\nu$  and
$m_{\mu\mu}^{(2)} \gg m_{ee}^{(2)} \gg m_{e\mu}^{(2)}$ in $ {\mathbb
M}_{\rm{2}}^\nu$, which is very similar to the case of solutions (C)
with a large $m_{ee}$ ($m_{\mu\mu}$) in $ {\mathbb
M}^{\rm{(1)}}_\nu$ (${\mathbb M}^{\rm{(2)}}_\nu$). Qualitatively,
these could form a different class of solutions compared to type (C)
with each of the sub-matrices here forming a quasi-degenerate pair
with small mixing. However, notice that   the total mixing is now
given by $ \tan 2 \theta \approx (x + x') /(\rho' - \rho)$ would
remain small as $x,x' \ll \rho, \rho'$ unless $\rho = \rho'$.  With
this \textit{additional} condition, this class of solutions again
falls in to the class (C) \textit{i.e}, $m^{(2)}_{ee} =
m^{(1)}_{\mu\mu}$ or $m^{(2)}_{\mu\mu} = m^{(1)}_{ee}$. However, to
distinguish from the solutions in Eq.~(\ref{massflsmallsum}), we
will call the class of solutions represented by Eq.~(\ref{2dsumdeg})
as type (C1)~\footnote{Solutions with three large entries in each
sub-matrix violate the small mixing assumption.}.

In summary, the sum of two mass matrices with small mixing angles
would naturally lead to a degenerate spectrum with maximal/large
mixing provided we insist there are no cancellations of the large
eigenvalues of the individual sub-matrices. The individual
sub-matrices could be {\sl{(a)}} ordered as hierarchial +
inverse-hierarchial with small mixing (solutions of type (C)) or
{\sl{(b)}} be quasi-degenerate themselves but with small mixing
(C1). However as we have seen, solutions of the type (C1) require
further precise cancellation in the differences of their large
diagonal elements. For this reason, we consider solutions of the
type (C) \textit{i.e}, matrices as parameterised in
Eq.~(\ref{massflsmallsum}) to be the most natural. Thus  to convert
\textit{one} small mixing angle  in two matrices to \textit{one}
maximal mixing in the total matrix, we would require  a
\textit{pair} of (quasi)-degenerate eigenvalues with the same CP
parities, ordered oppositely in the sub-matrices. This count would
be useful when we extend this \textit{degeneracy induced large
mixing} to three generations.

\subsection{Decomposition of the Degenerate Spectrum}
In the previous section we have seen that a quasi-degenerate pattern
naturally emerges if two mass matrices of small mixing are added and
we demand large mixing in the total mass matrix. One can instead
reverse the argument and might say that the quasi-degenerate
spectrum with large mixing can be decomposed in to two matrices with
small mixing.  In fact, the decomposition of the quasi-degenerate spectrum in
to two matrices is more generic and is independent of the mixing present
in them. This can be easily be demonstrated by considering zeroth
order neutrino mass matrices in the flavour basis.   Let us denote
the neutrino mass matrix in the flavour basis by
 \bea
{\mathbb M}_{\nu} ={\mathbb U}_{PMNS} {\mathbb {M}}_{diag} {\mathbb
U}_{PMNS}^\dagger~, \label{flmass} \eea where
${\mathbb{U}}_{PMNS}={\mathbb{U}}_l^\dagger {\mathbb{U}}_\nu $ is
the Pontecorvo-Maki-Nakagawa-Sakata (PMNS) unitary leptonic mixing
matrix. ${\mathbb{U}}_{PMNS}={\mathbb{U}}_\nu$ in a basis in which
charged lepton mass matrix is diagonal, {\textit{i.e}},
${\mathbb{U}}_l ={\mathbb{I}}_{n \times n}$.  In
Table~\ref{twoflavor},  we have listed the zeroth order mass
matrices  for hierarchal, inverse-hierarchal and degenerate spectra
for the case of  small mixing and maximal mixing. In writing down
these textures, we have followed Altarelli and
Feruglio~\cite{afreview} method, where each of these (zeroth order)
mass matrices  has to be multiplied by a mass scale $m$ representing
the heaviest eigenvalue of the mass matrix. From
Table~\ref{twoflavor} we can see that, as we go along each column,
the degenerate mass matrices $\mathbb{C}_i$ can be expressed as a
sum of hierarchal, $\mathbb{A}$ and inverse hierarchal, $\mathbb{B}$
matrices.  For example, $\mathbb{C}_0  = \mathbb{A} + \mathbb{B}$,
$\mathbb{C}_1 = \mathbb{A} - \mathbb{B}$, $\mathbb{C}_2 = \mathbb{B}
- \mathbb{A}$. Note that the mass scale $m$ multiplying $
\mathbb{C}_i$ now multiplies both $\mathbb{A}$ and $\mathbb{B}$.
These equations hold irrespective of the mixing being small or
maximal. Thus every degenerate mass matrix can be expressed a sum
(or difference) of a hierarchial and inverse-hierarchial mass
matrices, but  with common mass scale given by the degenerate mass
$m$, which is an obvious observation if one just sees the diagonal
eigenvalues of each  mass matrix in the first column.

%--------------------------------
%
%             Table
%
%--------------------------------
\begin{table}[h]
\begin{center}
\begin{tabular}{| l |  c  c | }
\hline
&& \\
\hspace{0.5cm}{\rm {Mixing $\Rightarrow$}}\hspace{0.5cm} &
\hspace{0.5cm}{\rm {Small}}\hspace{0.5cm} & \hspace{0.5cm}{\rm
{Maximal}}\hspace{0.5cm}
\\
\hspace{0.5cm}{\rm {$\mathbb{M}_{diag}$}}\hspace{0.5cm}& ${\mathbb{X}}_{\ep}$ & ${\mathbb{X}}_{M}$\\
&&\\
        \hline
Hierarchial &&\\
%        &&\\
        ${\mathbb{A}}$: Diag[0,1] & {\tiny{$\begin{pmatrix} 0 & \epsilon \\ \epsilon &
        1
\end{pmatrix}$}}
& {\tiny{$ \begin{pmatrix} 1/2 &1/2\\ 1/2&1/2
\end{pmatrix} $}}\\
\hline
Inverse hierarchial &&\\
${\mathbb{B}}$: Diag[1,0] &
{\tiny{$\begin{pmatrix} 1 & -\epsilon \\
 -\epsilon  &  0 \\
\end{pmatrix}$}} &
{\tiny{$ \begin{pmatrix} 1/2& -1/2  \\ -1/2& 1/2
\end{pmatrix} $}}
\\ \hline
Degenerate &&\\
${\mathbb{C}}_0$: Diag[1,1] & {\tiny{$\begin{pmatrix} 1 & 0 \\
0& 1
\end{pmatrix}$ }}& {\tiny{$\begin{pmatrix} 1 & 0 \\
 0 & 1
\end{pmatrix}$ }}
        \\
%\hline
       ${\mathbb{C}}_1$: Diag[-1,1] &{\tiny{ $\begin{pmatrix} -1 & 2 \epsilon_{} \\
2 \epsilon_{} & 1 \\
\end{pmatrix}$ }}& {\tiny{$\begin{pmatrix} 0 & 1  \\
1 & 0 \end{pmatrix}$ }} \\
%\hline
       ${\mathbb{C}}_2$: Diag[1,-1] &{\tiny{ $\begin{pmatrix} 1 & -2 \epsilon_{}\\
 -2 \epsilon_{} & -1 \\
%0& 2\epsilon_{23} & 1
\end{pmatrix}$ }}& {\tiny{$\begin{pmatrix} 0 & -1\\
-1 & 0
\end{pmatrix}$ }}\\
\hline
\end{tabular}
\caption[]{\footnotesize{Zeroth order textures  for small and
maximal  mixing (setting $m_1$ and $m_2$ as dimensionless quantities
which are either zero or one depending on the different cases
listed) for the two-generation case.} } \label{twoflavor}
\end{center}
\end{table}
%-----------------------------------------------------------

%-----------------------------------------------------------

Let us now turn to the question of mixing for the degenerate cases
mentioned above. The mixing in  $\mathbb{C}_0  =  \mathbb{A} +
\mathbb{B}$ in undetermined as it is proportional to the identity
matrix. This is also the exact degeneracy limit. This situation
arises if the mixing angles of  $\mathbb{A}$ and $\mathbb{B}$ are
not only small, but are also equal.  On the other hand, the mixing
in  $\mathbb{C}_1 = \mathbb{A} - \mathbb{B}$ can be maximal again as
we explained above in the previous section.  The mixing in
$\mathbb{C}_1$ and $\mathbb{C}_2$ will remain small as they have
opposite CP parities. An important exception to generate large
mixing in terms of small mixing angles through quasi-degeneracy is
the pseudo-Dirac pattern. The pseudo-Dirac pair can come as a sum
(difference) of two sub-matrices both with maximal mixing, one
hierarchal and the other inverse-hierarchal.  This is clearly
evident from the last column of Table~\ref{twoflavor}. We see  the
pseudo-Dirac pairs $\mathbb{C}_1 = \mathbb{A} -\mathbb{ B}$ and
$\mathbb{C}_2 = \mathbb{B}-\mathbb{A}$ with both $\mathbb{A}$ and
$\mathbb{B}$ containing maximal mixing.

 The decomposition of quasi-degenerate spectra can easily be incorporated within models
 of neutrino masses. For example, in the Type I seesaw mechanism (with two generations)
 the mass matrix is given by %
\bea -{\mathbb M}^{\rm{I}}_\nu &=& v^2 \begin{pmatrix}
h_{ee}^{D}  & h_{\mu e}^{D}  \\
h_{e\mu}^{D} & h_{\mu\mu} ^{D}
 \end{pmatrix}  \begin{pmatrix}
1/M_{R1}  &  0  \\
0  & 1/M_{R2}
 \end{pmatrix}
 \begin{pmatrix}
h_{ee}^{D}  & h_{e \mu}^{D}  \\
h_{\mu e}^{D} & h_{\mu\mu} ^{D}
 \end{pmatrix}
\nonumber \\
& = &m_1 \left( \begin{array}{cc}
(h_{ee}^D)^2 & h_{ee}^D h_{e \mu}^D \\
h_{ee}^D h_{e \mu}^D & (h_{e \mu}^D)^2 \end{array} \right)
+ m_2 \left( \begin{array}{cc}
(h_{\mu e}^D)^2 & h_{\mu e}^D h_{\mu \mu}^D \\
h_{\mu e}^D h_{\mu \mu}^D & (h_{\mu \mu}^D)^2 \end{array} \right),
\label{typeIdecomp}
\end{eqnarray}
where $m_1$ and $m_2$ are given as $v^2/(M_{R_1})$ and
$v^2/(M_{R_2})$ respectively.  Each of these sub-matrices is result
of a seesaw mechanism with one right-handed neutrino. Comparing the
above with Eq.~(\ref{massflsmallsum}), we can determine the
parameter regions required for quasi-degeneracy and large mixing.
For, $M_{R_1} = M_{R_2}$, we see that for the Yukawa parameters,
there are two choices where the mixing in the  sub-matrices is small
\begin{eqnarray}
h_{e \mu}^D ~\sim ~\mathcal{O}(1)~, \;\;
 h_{ee}^D ~\sim~ x~, \;\;
  h^D_{\mu \mu} ~\sim ~ -y~, \;\;
   h_{ \mu e }^D ~\sim ~\mathcal{O}(1)~, ~~~ \mbox{or}  \nonumber \\
h_{e e}^D ~\sim ~\mathcal{O}(1)~, \;\; h_{e \mu}^D ~\sim~ -x~, \;\;
h^D_{\mu e} ~\sim ~ y~, \;\;   h_{ \mu \mu }^D ~\sim
~\mathcal{O}(1)~.
\end{eqnarray}
 Thus each right-handed neutrino couples with the Standard Model neutrinos  with  only small mixing
 angles, but total mass matrix  ensures maximal mixing angles for the above choice of parameters.
 These conditions are already known in the literature for some time~\cite{afreview}. So this
is an alternative approach of arriving at these conditions.  The
interesting aspect of Type I seesaw mechanism is that the
decomposition at the neutrino mass  matrix level can be realised at
the Lagrangian level in terms of independent parameters with
`independent mass' scales for the the individual sub-matrices,  for
instance the sub-matrices have mass scales $v^2/M_{R_1}$ and
$v^2/M_{R_2} $. Such a realisation might not be possible in other
models for degenerate neutrinos like in Type II seesaw mechanism. A
further  interesting possibility would be to consider the case when
there are two independent seesaw mechanisms at work.

\subsection{Left-Right Symmetric Model}
The simplest model where the above mechanism can be realised is the
LRS model. In the recent years, this model has been thoroughly
analyzed for its duality properties~\cite{akhmedov}. The LRS model
naturally contains both Type I and Type II seesaw contributions,
which can be thought of as two sub-matrices discussed above. Further
more, these models are characterized by a common Yukawa coupling to
both the left-handed and right-handed Majorana mass matrices
\begin{equation}
\mathcal{L}_{M} = - {f \over 2} \left( \overline{\nu_L^c} \nu_L
\Delta^0_L + \overline{\nu_R^{c}}  \nu_R \Delta^0_R \right) + h.c.~,
\end{equation}
where $\Delta_{L(R)}$ is the triplet  Higgs field whose neutral
component attains a vacuum expectation value (\textit{vev}) giving
rise to the Majorana mass to the left (right) handed neutrino
fields.  In addition the Dirac neutrino Yukawa coupling is also
present
\begin{equation}
\mathcal{L}_{D} = -  Y \overline{\nu}_L \nu_R  \phi^0 + h.c.
\end{equation}
In the limit where $v_{R} \gg v$, the Type I seesaw mechanism
becomes operative and the total neutrino mass matrix is now given as
\begin{equation}
\label{type1plus2} {\mathbb M}_{\nu} = f v_L  - {v^2 \over v_R} Y
f^{-1} Y^T~.
\end{equation}
Along the lines of the discussion we had for the two-generation
case, Eq.~(\ref{massflsmallsum}), we can assume the  contribution
(first term on the RHS of Eq.~(\ref{type1plus2})), due to Type II to
be hierarchial with small mixing and second part due to the Type I
contribution  inverse hierarchial with small mixing. The appropriate
choice of the Yukawa textures in this case are  as follows
\begin{equation}
f = \left(\begin{array}{cc}
0 & x \\
x & 1 \\ \end{array} \right)  \;\;,\;\;
Y = \left(\begin{array}{cc}
1 & y \\
y & 0 \\ \end{array} \right)~.
\end{equation}
With this choice the total mass matrix takes the form
\begin{eqnarray}
{\mathbb M}_{\nu} &=& \left(\begin{array}{cc}
0 & m_1 x \\
m_1 x  & m_1  \\ \end{array} \right)  + { m_2 \over x^2}
 \left(\begin{array}{cc}
1 - 2 x y  & y(1 - xy)  \\
y(1-xy)  & y^2 \\ \end{array} \right) \nonumber \\
&=&
{1 \over x^2} \left(\begin{array}{cc}
m_2 (1 - 2 xy) & m_1 x^3 +  m_2 y(1 - xy)  \\
m_1 x^3 + m_2 y(1-xy)  & m_1 x^2 + m_2 y^2 \\ \end{array} \right)~,
\end{eqnarray}
where $m_1 = v_L$ and $m_2 = v^2/v_R$. The mixing angle in the above
mass matrix is given by
\begin{equation}
\tan 2 \theta = { 2 [ m_1 x^3 +  m_2 y(1 - xy) ] \over [m_1 x^2 +
m_2 y^2] - m_2 ( 1 - 2 xy)}~.
\end{equation}
 \noindent
 From the above it is clear that the degeneracy requirement $m_1 x^2 \approx m_2$
  automatically leads to large mixing, $\tan 2 \theta\sim {\cal O}(\frac{1}{2 y})$.
   A rough idea of how stable this mixing would be under radiative corrections
 can be obtained by considering the modification of the neutrino mass matrix below the seesaw scale.
 The modification is set by the matrix ${\mathbb{P}}  =\mbox{Diag}\{1,1+\delta_\mu\} $  and is given as
 ${\mathbb{M}}_\nu = {\mathbb{P}} {\mathbb{M}}_\nu {\mathbb{P}} $.  In this case, the mixing angle now takes the form
\begin{equation}
\tan 2 \theta = { 2[ m_1 x^3 +  m_2 y(1 - xy) ] ( 1 + \delta_\mu)
\over (1 + \delta_\mu)^2 [m_1 x^2 + m_2 y^2] - m_2 ( 1 - 2 xy)}~,
\end{equation}
 \noindent
 where $\delta_\mu = c ~h_\mu^2/(16 \pi^2) \log(M_X/M_W) $ specifies the size of the radiative corrections
  induced by the Yukawa coupling of the $\mu$, $h_\mu$. Here   $c$
  is a constant  depending on whether the theory is supersymmetric or not
 and $M_X$ is the high scale just below the seesaw scale~\cite{antusch}. The condition for large mixing case now gets
modified as  $[m_1 x^2 + m_2 y^2] (1 + \delta_\mu)^2  \approx m_2 $.
Note that this same condition is also required to keep the
degeneracy stable even after radiative corrections.  Of course, the
splitting of the degeneracy can come from the radiative effects. A
more detailed analysis of radiative corrections will be presented
elsewhere.

 \section{Extension to Three Generations}
 \label{secthree}
 Let us extend the analysis of the previous section to the case of three generations.
 Here we will consider two cases - {\sl{(a)}} Case I: two seesaw mechanisms or
 two sources of neutrino masses (Sec.~\ref{twsm}), and {\sl{(b)}} Case II: three seesaw mechanisms or
 three sources of neutrino masses (Sec.~\ref{thsm}).

 \subsection{Case I: Two Seesaw Mechanisms}
 \label{twsm}
 As before,  let us consider two $3\times 3$ mass matrices each with a small mixing angle and one
large eigenvalue, ${\mathbb M}_\nu = {\mathbb M}_{\nu}^{(1)} + {\mathbb M}_{\nu}^{(2)}$.
  Instead of representing them as general mass matrices  as we have done for the case
  of two generations, we will represent them by using
\begin{equation}
\mathbb{M}_\nu^{(i)} = [{\mathbb{U}}_{\mbox{mix}}^{(i)}]^T \cdot
\mbox{Diag}[\mathbb{M}_\nu^{(i)}] \cdot
{\mathbb{U}}_{\mbox{mix}}^{(i)}~,
\end{equation}
where $\mbox{Diag}[\mathbb{M}_\nu^{(i)}] = \mbox{Diag}[\{ m_1^{(1)},
m_2^{(2)}, m_3^{(3)} \}]$, the eigenvalues of the mass matrices and
${\mathbb{U}}_{\mbox{mix}}^{(i)}$ represents the mixing present in
each of the mass matrices, with $i = 1,2$. Given that the mixing
angles in $\mathbb{M}_\nu^{(i)}$ are small, we can expand
${\mathbb{U}}_{\mbox{mix}}^{(i)}$ in terms of small parameters
$\cos\theta_{m}^{(i)} \approx 1 $, $ \sin\theta_{m}^{(i)} \approx
\epsilon^{(i)}_m$, where $m = \{12,23,13\}$  labels the three
angles. The total mass matrix now takes the form
\begin{equation}
\mathbb{M}_\nu = \left( \begin{array}{ccc}
m_1^{(1)} + m_1^{(2)} & (m_2^{(1)} - m_1^{(1)} ) \epsilon^{(1)}_{12} + (m_2^{(2)} - m_1^{(2)} ) \epsilon^{(2)}_{12} &
(m_3^{(1)} - m_1^{(1)} ) \epsilon^{(1)}_{13} + (m_2^{(2)} - m_1^{(2)} ) \epsilon^{(2)}_{13} \\
*& m_2^{(1)} + m_2^{(2)} & (m_3^{(1)} - m_2^{(1)} ) \epsilon^{(1)}_{23} + (m_3^{(2)} - m_2^{(2)} ) \epsilon^{(2)}_{23}\\
* &* & m_3^{(1)} + m_3^{(2)} \end{array} \right),
\end{equation}
where the symmetric elements of the matrix have been represented by $*$.
We can determine the mixing present in the total mass matrix by diagonalising
the above matrix.  We have
\begin{equation}
\mathbb{M}'_\nu = {\mathbb{U}}_{23}^T \mathbb{M}_\nu
{\mathbb{U}}_{23}~,
\end{equation}
where $${\mathbb{U}}_{23} = \left( \begin{array}{ccc}
1&0&0 \\
0& \cos\theta_{23} & \sin \theta_{23} \\
0& -\sin\theta_{23} & \cos \theta_{23}
\end{array} \right)~, $$ with
\begin{equation}
\tan 2 \theta_{23} = 2 ~{ (m_3^{(1)} - m_2^{(1)})
\epsilon^{(1)}_{23} +( m_3^{(2)} - m_2^{(2)} )\epsilon^{(2)}_{23}
\over m_3^{(1)} + m_3^{(2)} - m_2^{(1)} - m_2^{(2)} }
~.\end{equation} For this mixing to be maximal the condition would
be $(m_3^{(1)} - m_2^{(1)}) =  - (m_3^{(2)} - m_2^{(2)})$. This
condition is similar to the one we have seen earlier for the two
generation case and as argued in that case, the only natural
solution is to have $m_3^{(1)}  =   m_2^{(2)}$ with $ m_2^{(1)},
m_3^{(2)}$ negligible or  $m_2^{(1)}  =  m_3^{(2)}$ with~\footnote{
More precisely, we should have $m_3^{(1)} -  m_2^{(2)}\approx {\cal
O} (m_3^{(1)}(\epsilon_{23}^1-\epsilon_{23}^2)) $and
$m_{2}^{(1)},m_{2}^{(2)}$ much smaller compared to them.}
$m_3^{(1)}, m_2^{(2)}$ negligible. We now proceed to show that  if
we accept either of these two solutions, it would not be possible to
have one another large mixing angle in $\mathbb{M}_\nu$,  if they
have to satisfy the naturalness criteria that the large eigenvalues
of the individual matrices should not cancel in the total mass
matrix. Defining
\begin{equation}
{\mathbb{U}}_{13} = \left( \begin{array}{ccc}
\cos \theta_{13} &0& \sin \theta_{13} \\
0 & 1 & 0 \\
- \sin \theta_{13} & 0 & \cos \theta_{13} \end{array} \right)~,
\end{equation}
we have
\begin{equation}
\mathbb{M}^{''}_\nu = {\mathbb{U}}_{13}^T \mathbb{M}'_\nu
{\mathbb{U}}_{13}~.
\end{equation}
 $\tan 2 \theta_{23}$  in the limit where the solution for maximal mixing of the
 $23$ angle, $m_3^{(1)}  =  m_2^{(2)} = \bar{m}$ with $m_2^{(1)}, m_3^{(2)} \sim 0$
 is taken is given by
\begin{equation}
\tan 2\theta_{13} \approx { m_1^{(1)} ( \epsilon^{(1)}_{12} +
\epsilon^{(1)}_{13} ) - \bar{m} ( \epsilon^{(1)}_{13} + \epsilon^{(2)}_{12}) + m_1^{(2)}
 (\epsilon^{(2)}_{12} + \epsilon^{(2)}_{13} ) \over \sqrt{2} (m_1^{(1)} + m_1^{(2)}
 - \bar{m} ( 1 + \epsilon^{(1)}_{23} - \epsilon^{(2)}_{23}) )}~.
  \end{equation}
From the above we realize the following conditions for {\sl{(a)}}
small mixing : $|m_1^{(1)} + m_1^{(2)} - \bar{m}| \gg 0$, and
{\sl{(b)}} maximal mixing : $|m_1^{(1)} + m_1^{(2)} - \bar{m}| = 0$.
Finally, defining
\begin{equation}
{\mathbb{U}}_{12} =  \left( \begin{array}{ccc}
\cos \theta_{12} & \sin \theta_{12} &0 \\
- \sin \theta_{12} &  \cos \theta_{12}&0 \\
0 & 0& 1
 \end{array} \right)~,
\end{equation}
we have
\begin{equation}
\mathbb{M}^{'''}_\nu = {\mathbb{U}}_{12}^T \mathbb{M}^{''}_\nu
{\mathbb{U}}_{12}~.
\end{equation}
$\tan 2\theta_{12}$ has the following form in the limiting case when
$\theta_{13}$ is very small
\begin{equation}
\tan 2 \theta_{12} \approx {m_1^{(1)} (\epsilon_{12}^{(1)} - \epsilon_{13}^{(1)}) + \bar{m} (\epsilon_{13}^{(1)} - \epsilon_{12}^{(2)}) +
 m_1^{(1)} (\epsilon^{(2)}_{12} -\epsilon^{(2)}_ {13}) \over \sqrt{2} (m_1^{(1)} + m_1^{(2)} - \bar{m}(1 -  \epsilon_{23}^{(1)} + \epsilon_{23}^{(2)}))} + \mathcal{O}(
 \theta_{13})~. \end{equation} From the above we see that the conditions for the
mixing are the same for both $\theta_{12}$ and $\theta_{13}$ in this
limit. Thus either both become maximal/large or both remain small.
Finally, in the limit of maximal $\theta_{13}$ mixing, the
expression for $\tan 2 \theta_{12}$ becomes
\begin{equation}
{(m_1^{(1)} (\epsilon_{12}^{(1)} - \epsilon_{13}^{(1)}) + \bar{m} (\epsilon_{13}^{(1)} -
 \epsilon_{12}^{(2)}) + m_1^{(2)} (\epsilon_{12}^{(2)} - \epsilon_{13}^{(2)})) \over
m_1^{(1)} + m_1^{(2)}  +  \bar{m} (-1 + 3 \epsilon_{23}^{(1)} - 3 \epsilon^{(2)}_{23}) +
 \sqrt{2} (m_1^{(1)} (\epsilon_{12}^{(1)} + \epsilon_{13}^{(1)}) - \bar{m} (\epsilon_{13}^{(1)}
 + \epsilon_{12}^{(2)}) + m_1^{(2)} (\epsilon^{(2)}_{12} + \epsilon_{13}^{(2)}))
 }~,
 \end{equation}
which is also automatically maximal/large within the  small
$\epsilon_{ij}^{(k)}$ limit. Before we proceed, a few comments are
in order regarding the ordering of the eigenvalues. In the case
where there is only one maximal/large mixing,  the sub matrices can
have hierarchal and inverse-hierarchal patterns, with the hierarchal
sub-matrix  containing one large eigenvalue and the
inverse-hierarchal containing two large eigenvalues. The only
condition is on their CP parties;  the eigenvalues taking part in
the enhancement of the mixing should have the same CP parities.  The
list of possible forms the sub-matrices can take is discussed in the
subsection~\ref{decomdeg} where decomposition of the degenerate
spectrum is considered in three generation case.

On the other hand, for the case with all the three large/maximal mixing case, as per
our arguments  earlier, \textit{i.e,} the large eigenvalues of the individual matrices
should not cancel in the total matrix, the present solution necessarily favours an
\textit{alternating} pattern for the eigenvalues for the individual
mass matrices~\footnote{The zeroth order textures for alternating
pattern of neutrino mass matrices are given in
Table~\ref{threeflavornewalt}.}
\begin{eqnarray} \label{alttex}
\mbox{Diag}[\mathbb{M}_{\nu}^{(1)}] =  \mbox{Diag}[\{m_1^{(1)},0,m_3^{(1)}\}] &,& \mbox{Diag}[ \mathbb{M}_{\nu}^{(2)} =  \mbox{Diag}[\{0,m_2^{(2)},0 \} ]\nonumber \\
\mbox{Diag} [\mathbb{M}_{\nu}^{(1)} ] =
\mbox{Diag}[\{0,m_2^{(1)},0\}] &,& \mbox{Diag}
[\mathbb{M}_{\nu}^{(2)} ] =  \mbox{Diag}[\{m_1^{(2)},0,m_3^{(2)}
\}]~.
\end{eqnarray}
In this case, the mixing pattern corresponds to the
 \textit{truly maximal mixing} matrix of Cabibbo and
Wolfenstein~\cite{cabwolf} along with the degeneracy
condition $m_1 \approx m_2 \approx m_3$. In the recent years,  the truly maximal
mixing matrix has been achieved from $A_4$ symmetry by Ma and
Rajasekaran also for degenerate case~\cite{marajasekaran}. As it stands
this matrix is not phenomenologically viable as all the three mixing
angles it predicts are large. However there could be other
corrections to this mass matrix depending on the model which would
rectify this situation~\cite{vallemababu} and make the mass matrix
phenomenologically viable.

\subsubsection{Two Equivalent Textures}
\label{tet} From our arguments above, it appears that we can
generate only one large mixing angle in the case when there are only
two sub-matrices, because of the important constraint that the third
 mixing angle ($\theta_{13}$) must not be large~\footnote{This would be
case in models where there are no large radiative corrections effecting
the mixing angles strongly.}. Given that we can
 only generate one large mixing from the small mixing using the
 degenerate conditions, we will have to assume that at least one of the
 sub-matrices has intrinsically one maximal/large mixing angle. However, the presence of this
 mixing should not disturb the
  smallness of  $\theta_{13}$ angle in the total mass matrix. In the following, we
  will consider one of the sub-matrices to have pseudo-Dirac structure and other one
   to have one large eigenvalue and all the three mixing angles small. This is because
   the pseudo-Dirac structure not only gives maximal mixing but also has the eigenvalues
   with opposite CP parities.
\begin{equation}
{\mathbb{M}}_\nu = m_1 \left( \begin{array}{ccc} x^2 & x & y^2 \\
x & 0 & 1 \\
y^2 & 1 & 0 \end{array} \right) + m_2 \left( \begin{array}{ccc}
1 & z & t^3 \\
z & z^3 &  t^3 \\
t^3 & t^3 & z^3 \end{array} \right)~,
\label{twoequiv}
\end{equation}
where $x, y, z, t$ are small entries compared to $m_1, m_2$. We will
diagonalise this matrix  in the following manner.  Rotating by
${\mathbb{O}}_{23}$ on both sides, we have
\begin{equation}
\label{23rotation} {\mathbb{O}}_{23}^T {\mathbb{M}}_{\nu}
{\mathbb{O}}_{23} = \left(
\begin{array}{ccc}
m_2 + m_1 x^2  &  m_1 x \cos \theta_{23} + \widetilde{m}_{12} & -m_1 x \sin \theta_{23} + \widetilde{m}_{13} \\
 m_1 x \cos \theta_{23} + \widetilde{m}_{12} & m_1   \sin 2 \theta_{23} + \widetilde{m}_{22} & 0 \\
- m_1 x \sin \theta_{23} + \widetilde{m}_{13}  &  0 &  - m_1  \sin 2
\theta_{23} + \widetilde{m}_{33}  \end{array} \right)~,
 \end{equation}
 where ${\mathbb{O}}_{23}$ is defined as
 \begin{equation}
 \label{23mix}
 {\mathbb{O}}_{23} = \left( \begin{array}{ccc}
 1 & 0 & 0 \\
 0& \cos \theta_{23} & - \sin \theta_{23} \\
 0 & \sin \theta_{23} & \cos \theta_{23}  \end{array} \right)~,
 \end{equation}
 with the angle $\theta_{23}$ given by
 \begin{equation}
 \theta_{23} = {1 \over 2} \tan^{-1} \left[
  {2 (m_1 + m_2 t^3) \over ( m_2 z^3 - m_2 z^3)  } \right]  =  {\pi \over
  4}~.
 \end{equation}
 The explicit forms for $\widetilde{m}_{ij}$ can be easily deduced.  A crucial point to note is that the diagonal elements
 of the matrix in Eq.~(\ref{23rotation}) carry opposite sign
  for the dominant element ($m_1$).  This would have the
 consequence of keeping the $13$ mixing small,
 while making  $23$ mixing large, when the degeneracy condition
  $m_1 \approx  m_2 \approx m $ is imposed. The total mixing matrix
  is given by ${\mathbb{O}} = {\mathbb{O}}_{12} {\mathbb{O}}_{13} {\mathbb{O}}_{23}$ with
  the angles $\theta_{13}$ and $\theta_{12}$ defined as
  \begin{eqnarray}
  \theta_{13} & =& {1 \over 2} \tan^{-1} \left[
  {2 ( -m_1 x \sin \theta_{23} + \widetilde{m}_{13})
  \over    - m_1  \sin 2 \theta_{23} + \widetilde{m}_{33} -   m_1
  x^2 - {m}_{2} } \right]~, \nonumber \\
  \theta_{12} &=&   {1 \over 2} \tan^{-1} \left[
  { 2 \widetilde{m}'_{12} \over \widetilde{m}'_{22} - \widetilde{m}'_{11} }
  \right]~,
   \end{eqnarray}
 where the explicit form of $\widetilde{m}'_{ij}$ can easily be deduced. From the above, we can see that the degeneracy induced large mixing
 mechanism works for the $12$ mixing, while it does not generate large (maximal) mixing for the $13$ mixing angle.  This is due to the choice
 of having $\tau \tau$ element with opposite sign (loosely speaking CP parity) compared to the $\mu\mu$ element.

The above Yukawa matrices can be easily incorporated in the LRS
model by choosing $f$ and $Y$ of Eq.~(\ref{type1plus2})  (at the leading order) as
\begin{equation}
f = \left( \begin{array}{ccc} x^2 & x & y^2 \\
x & 0 & 1 \\
y^2 & 1 & 0 \end{array} \right)  \;\;\;  Y =  \left( \begin{array}{ccc} 1 & 0 & 0 \\
0 & 0 & 0 \\
0 & 0 & 0 \end{array} \right)~.\end{equation} Notice that it
reproduces the Eq.~(\ref{twoequiv}) at the zeroth order. From the
discussion in the previous section, we also know that the
pseudo-Dirac mass matrix can be decomposed in to maximally mixing
sub-matrices.  Thus another texture which could equally give the
same results is given by
\begin{equation}
{\mathbb{M}}_\nu = m_1 \left( \begin{array}{ccc} x^2 & x & y^2 \\
x & {1 \over 2} & {1 \over 2}\\
y^2 & {1 \over 2} & {1 \over 2}  \end{array} \right) + m_2 \left( \begin{array}{ccc}
1 & z & t^3 \\
z & {1 \over 2}  &  - {1 \over 2} \\
t^3 & -{1 \over 2} & {1 \over 2}  \end{array} \right)~,
\end{equation}
The first of the matrices has only one large eigenvalue in a
hierarchial pattern with maximal mixing, whereas the second one has
two large eigenvalues with one
 maximal mixing and two small mixings with inverted hierarchy.  Lets emphasize once more that
one needs opposite eigenvalues $m_1\approx -m_2$ to obtain the large atmospheric mixing in this case.

\subsubsection{Decomposition of the Degenerate Spectrum}
\label{decomdeg} For three generations the decomposition of the
degenerate spectrum in to hierarchal and inverse-hierarchal mass
patterns is straight forward. In Table~\ref{threeflavor}, we present
the zeroth order mass matrices for the three generation case.
%--------------------------------
%
%             Table
%
%--------------------------------
\begin{table}[h]
\begin{center}
\begin{tabular}{| l | c  c  c  c |}
\hline
&&& &\\
\hspace{0.5cm}{\rm {Mixing $\Rightarrow$}}\hspace{0.5cm} &
\hspace{0.5cm}{\rm {Small}}\hspace{0.5cm} & \hspace{0.5cm}{\rm
{Single maximal}} \hspace{0.5cm} & \hspace{0.5cm} {\rm
{Bimaximal}}\hspace{0.5cm} & \hspace{0.5cm} {\rm
{Tribimaximal}}\hspace{0.5cm}
\\

\hspace{0.5cm}{\rm{$\mathbb{M}_{diag}$}}\hspace{0.5cm} & ${\mathbb{X}}_\ep$ & ${\mathbb{X}}_{SM}$ & ${\mathbb{X}}_{BM}$ & ${\mathbb{X}}_{TBM}$\\
&&&&\\
        \hline
Hierarchial &&& &\\
%        &&\\
        ${\mathbb{A}}$: Diag[0,0,1] & {\tiny{$\begin{pmatrix} 0 &0& \epsilon_{13} \\ 0&0& \epsilon_{23} \\
\epsilon_{13} & \epsilon_{23}& 1
\end{pmatrix}$}}
&
{\tiny{$ \begin{pmatrix} 0 &0& 0 \\ 0&\frac{1}{2} & \frac{1}{2}  \\
0 & \frac{1}{2}  & \frac{1}{2} \end{pmatrix} $}}  &{\tiny{$ \begin{pmatrix} 0 &0& 0 \\ 0& \frac{1}{2} & \frac{1}{2}  \\
0 &  \frac{1}{2} & \frac{1}{2}
\end{pmatrix} $}} & {\tiny{$
\begin{pmatrix}
 0 & 0 & 0
 \\  0 & \frac{1}{2} &  \frac{1}{2} \\
 0 &\frac{1}{2} & \frac{1}{2} \end{pmatrix}
  $}}    \\
\hline
Inverse hierarchial &&&&\\
       $ {\mathbb{B}}_1$: Diag[1,-1,0]&
{\tiny{$\begin{pmatrix} 1 & -2 \epsilon_{12} & -\epsilon_{13} \\
 -2 \epsilon_{12} & -1 & \epsilon_{23} \\
-\epsilon_{13} & \epsilon_{23}& 0
\end{pmatrix}$}} &{\tiny{$\begin{pmatrix} 1 &0& 0 \\ 0&- \frac{1}{2} & \frac{1}{2} \\
0 & \frac{1}{2} & -\frac{1}{2}
\end{pmatrix}$}} &
{\tiny{$ \begin{pmatrix} 0 & -\frac{1}{\sqrt 2}
& \frac{1}{\sqrt 2}  \\ -\frac{1}{\sqrt 2} & 0 & 0 \\
\frac{1}{\sqrt 2} & 0 & 0
\end{pmatrix} $}}
& {\tiny{$ \begin{pmatrix}
 \frac{1}{3} & - \frac{2}{3} & \frac{2}{3}\\
 - \frac{2}{3}&  - \frac{1}{6}  & \frac{1}{6} \\
 \frac{2}{3}& \frac{1}{6} & -\frac{1}{6}
\end{pmatrix} $}}
\\
${\mathbb{B}}_2$: Diag[1,1,0] &
{\tiny{$\begin{pmatrix} 1 &0& -\epsilon_{13} \\
0&1& -\epsilon_{23} \\
-\epsilon_{13} & -\epsilon_{23}& 0
\end{pmatrix}$}} &
{\tiny{$ \begin{pmatrix} 1& 0 & 0 \\ 0& \frac{1}{2} & -\frac{1}{2}  \\
0 & - \frac{1}{2} & \frac{1}{2}
\end{pmatrix} $}}
&{\tiny{
$\begin{pmatrix} 1 &0& 0 \\ 0& \frac{1}{2}  & -\frac{1}{2}  \\
0 & -\frac{1}{2}  & \frac{1}{2}
\end{pmatrix}$}}&
{\tiny{$ \begin{pmatrix} 1 & 0 & 0\\
0 & \frac{1}{2} & -\frac{1}{2}\\
0 & -\frac{1}{2} & \frac{1}{2}
 % 0 &0& 0 \\ 0&1/2& -1/2 \\
%0 & -1/2& 1/2
\end{pmatrix} $}}
         \\
\hline
Degenerate  &&&&\\
${\mathbb{C}}_0$: Diag[1,1,1] & {\tiny{$\begin{pmatrix} 1 &0 & 0 \\
 0 & 1 & 0 \\
0& 0& 1
\end{pmatrix}$ }}& {\tiny{$\begin{pmatrix} 1 &0 & 0 \\
 0 & 1 & 0 \\
0& 0& 1
\end{pmatrix}$ }}&{\tiny{
$
\begin{pmatrix} 1 &0& 0 \\ 0&1& 0 \\
0 & 0& 1
\end{pmatrix}$}}&
{\tiny{$ \begin{pmatrix} 1 & 0 &
0\\
0 & 1 & 0
 \\
0 & 0
 & 1
\end{pmatrix} $}}
        \\
%\hline
       ${\mathbb{C}}_1$: Diag[-1,1,1] &{\tiny{ $\begin{pmatrix} -1 & 2 \epsilon_{12}& 2 \epsilon_{13} \\
 2 \epsilon_{12} & 1 & 0 \\
2 \epsilon_{13} & 0 & 1
\end{pmatrix}$ }}& {\tiny{
$ \begin{pmatrix} -1 &0& 0 \\ 0&1& 0 \\
0 & 0& 1
\end{pmatrix}
$ }} & {\tiny{$\begin{pmatrix} 0 & \frac{1}{\sqrt{2}} & -\frac{1}{\sqrt{2}} \\
\frac{1}{\sqrt 2}  & \frac{1}{2}  &  \frac{1}{2}  \\
- \frac{1}{\sqrt 2} &  \frac{1}{2}  & \frac{1}{2}
\end{pmatrix}$ }} &{\tiny{$ \begin{pmatrix}
- \frac{1}{3} & \frac{2}{3} & -\frac{2}{3}
 \\ \frac{2}{3}  & \frac{2}{3}  & \frac{1}{3} \\
- \frac{2}{3} & \frac{1}{3} & \frac{2}{3}
\end{pmatrix} $}}   \\
%\hline
       ${\mathbb{C}}_2$: Diag[1,-1,1] &{\tiny{ $\begin{pmatrix} 1 & -2 \epsilon_{12} & 0 \\
 -2\epsilon_{12} & -1 & 2 \epsilon_{23} \\
0& 2\epsilon_{23} & 1
\end{pmatrix}$ }}&{\tiny{
$\begin{pmatrix} 1 &0& 0 \\ 0&0& 1 \\
0 & 1& 0
\end{pmatrix}
$ }}& {\tiny{$\begin{pmatrix} 0 &
 -\frac{1}{\sqrt{2}} & \frac{1}{
\sqrt{2}} \\
-\frac{1}{\sqrt{2}} & \frac{1}{2} & \frac{1}{2} \\
\frac{1}{\sqrt{2}}& \frac{1}{2} &\frac{1}{2}
\end{pmatrix}$ }} & {\tiny{$ \begin{pmatrix}
\frac{1}{3} & -\frac{2}{3} & \frac{2}{3}
 \\- \frac{2}{3}  & \frac{1}{3}  & \frac{2}{3} \\
\frac{2}{3} & \frac{2}{3} & \frac{1}{3}
\end{pmatrix} $}}  \\
%\hline
      $ {\mathbb{C}}_3$: Diag[1,1,-1] &{\tiny{ $\begin{pmatrix} 1 & 0 & -2 \epsilon_{13}  \\
 0 & 1 & -2 \epsilon_{23} \\
-2 \epsilon_{13}& -2\epsilon_{23} & -1
\end{pmatrix}$}}
&{\tiny{
$\begin{pmatrix} 1 &0& 0 \\ 0&0& -1 \\
0 & -1& 0
\end{pmatrix}$ }}&
{\tiny{
$\begin{pmatrix} 1 &0& 0 \\ 0&0& -1 \\
0 & -1& 0
\end{pmatrix}$}}&
{\tiny{$ \begin{pmatrix}
1 & 0 & 0 \\ 0 & 0 & -1 \\
0 & -1 & 0
\end{pmatrix} $}}  \\
%&& \\
\hline
\end{tabular}
\caption[]{\footnotesize{Different standard textures (zeroth order)
for different combinations of mixings (setting $m_1$, $m_2$ and
$m_3$ as dimensionless quantities which are either zero or one
depending on the different cases listed) consistent with
data~\footnote{In Ref.~\onlinecite{afreview}, only two cases (single
and bimaximal mixing) were considered and they used ${\mathbb
M}_{\nu}={\mathbb U}_{PMNS}^\dagger {\mathbb M}_{diag} {\mathbb
U}_{PMNS}$, which is different from our definition
(Eq.~(\ref{flmass})). }.}} \label{threeflavor}
\end{center}
\end{table}
%-----------------------------------------------------------
 Note that the present
notation has been previously used  in the literature~\cite{afreview}
and all the matrices present in this table have been used {\bf previously}  to
describe the neutrino mass matrix at the zeroth order.  After adding
\textit{small} perturbations to these matrices they can explain the
neutrino data. However, as before we are interested in only
decomposing the degenerate mass matrix in terms of the hierarchal
and inverse-hierarchal mass matrices. As before,  from  each of the
columns,  we can see that each degenerate case can be constructed as
a sum of hierarchal and inverse hierarchal textures. For example,
$\mathbb{C}_0$ can be considered as
 $\mathbb{A} ~+ ~ \mathbb{B}_2$. Similarly, $\mathbb{C}_1$ can be
 considered  as $-\mathbb{B}_1 ~+~\mathbb{A}$ and so on.  And this
 is true as we go along each of the columns,  \textit{i.e} for all kinds of
 mixing angles.  This simple observation can be restated as \textit{every
 degenerate neutrino mass matrix can be thought of a sum of hierarchal
 and inverse hierarchal sub-mass matrices while the converse is not
 generally true}.
%--------------------------------
%
%             Table
%
%--------------------------------
\begin{table}[h]
\begin{center}
\begin{tabular}{| l | c  c c  c |}
\hline
&&& &\\
\hspace{0.5cm}{\rm {Mixing $\Rightarrow$}}\hspace{0.5cm} &
\hspace{0.5cm}{\rm {Small}}\hspace{0.5cm} & \hspace{0.5cm}{\rm
{Single maximal}} \hspace{0.5cm} & \hspace{0.5cm}{\rm
{Bimaximal}}\hspace{0.5cm} & \hspace{0.5cm}{\rm
{Tribimaximal}}\hspace{0.5cm}
\\
\hspace{0.5cm}{\rm {$\mathbb{M}_{diag}$}}\hspace{0.5cm} & ${\mathbb{X}}_\ep$ & ${\mathbb{X}}_{SM}$ & ${\mathbb{X}}_{BM}$ & ${\mathbb{X}}_{TBM}$\\
&&&&\\
        \hline
% &&& &\\
$\widetilde {\mathbb{A}}_1$: Diag[0,1,1] & {\tiny{$\begin{pmatrix}
0 & \ep _{12} & \epsilon_{13} \\
\ep_{12}& 1 & 0 \\
\epsilon_{13} & 0 & 1
\end{pmatrix}$}}
&
{\tiny{$ \begin{pmatrix} 0 & 0& 0 \\ 0& 1  & 0  \\
0 & 0 & 1 \end{pmatrix} $}}
  &{\tiny{$
  \begin{pmatrix} \frac{1}{2} & \frac{1}{2\sqrt{2}} & -\frac{1}{2\sqrt{2}}
   \\ \frac{1}{2\sqrt{2}} & \frac{3}{4} & \frac{1}{4}  \\
-\frac{1}{2\sqrt{2}} & \frac{1}{4} & \frac{3}{4}
\end{pmatrix} $}} & {\tiny{$
\begin{pmatrix}
  \frac{1}{3} & \frac{1}{3} & -\frac{1}{3}
 \\ \frac{1}{3} & \frac{5}{6} & \frac{1}{6}
  \\-\frac{1}{3} &  \frac{1}{6}& \frac{5}{6}  \end{pmatrix} $}}    \\
 $\widetilde {\mathbb{A}}_2$:
 Diag[0,1,-1] & {\tiny{$\begin{pmatrix}
 0 & \ep_{12} & -\epsilon_{13} \\ \ep_{12} & 1 & -2\epsilon_{23} \\
-\epsilon_{13} & -2\epsilon_{23}& -1
\end{pmatrix}$}}
&
{\tiny{$ \begin{pmatrix} 0 &0& 0 \\ 0& 0 & -1  \\
0 & -1 & 0 \end{pmatrix} $}}  &{\tiny{$ \begin{pmatrix} \frac{1}{2}
& \frac{1}{2\sqrt{2}} & -\frac{1}{2\sqrt{2}}
   \\ \frac{1}{2\sqrt{2}} & -\frac{1}{4} & -\frac{3}{4}  \\
-\frac{1}{2\sqrt{2}} & -\frac{3}{4} & -\frac{1}{4}
\end{pmatrix} $}} & {\tiny{$
\begin{pmatrix}
 \frac{1}{3}& \frac{1}{3} & -\frac{1}{3}
 \\ \frac{1}{3} & -\frac{1}{6}& -\frac{5}{6}  \\
 -\frac{1}{3} & -\frac{5}{6}& -\frac{1}{6} \end{pmatrix}
  $}}    \\
\hline
%IH &&&&\\
\ $\widetilde {\mathbb{B}}$: Diag[1,0,0] &
{\tiny{$\begin{pmatrix} 1 &-\ep_{12}& -\epsilon_{13} \\
-\ep_{12} & 0 & 0 \\
-\epsilon_{13} & 0 & 0
\end{pmatrix}$}} &
{\tiny{$ \begin{pmatrix} 1& 0 & 0 \\ 0& 0 & 0 \\
0 & 0 & 0
\end{pmatrix} $}}
&{\tiny{ $\begin{pmatrix} \frac{1}{2} & -\frac{1}{2\sqrt{2}} &
\frac{1}{2\sqrt{2}}
   \\ -\frac{1}{2\sqrt{2}} &\frac{1}{4} & -\frac{1}{4}  \\
\frac{1}{2\sqrt{2}} & -\frac{1}{4} & \frac{1}{4}
\end{pmatrix}$}}&
{\tiny{$ \begin{pmatrix} \frac{2}{3} & -\frac{1}{3} & \frac{1}{3} \\
-\frac{1}{3} & \frac{1}{6} & -\frac{1}{6}\\
\frac{1}{3} & -\frac{1}{6} & \frac{1}{6}
\end{pmatrix} $}}\\
\hline
\end{tabular}
\caption[]{\footnotesize{Novel textures (leading order) for
different mixing scenarios which by themselves need not be
consistent with data. These cases are useful when we consider adding
two different textures to obtain the degenerate cases. The labels
with tilde sign are new textures by taking into account the fact
that hierarchy or inverse hierarchy can appear in either 1-2 sector
or the 2-3 sector respectively. The standard textures considered
degeneracy in 1-2 sector and hierarchy or inverse hierarchy only in
the 2-3 sector.
 }} \label{threeflavornew}
\end{center}
\end{table}
%-----------------------------------------------------------
In three generations,  the above set of decomposition which is based
on neutrino data is not exhaustive.  This is essentially because the
constraints of the neutrino data are not on the individual
sub-matrices but on the total mass matrix.
 In such a case,  the normal and inverse hierarchial
sub matrices can take other possible forms ${\mathbb{A}}$ and
${\mathbb{B}}_i$ than those listed in Table~\ref{threeflavor}.
% We
%can interchange $\mathbb{A}$ and $\mathbb{B}$ types which are  given
%in Table~\ref{threeflavornew}.
From Table~\ref{threeflavornew}, it is easy to see that  the
combinations of $\widetilde{\mathbb{A}}_i$ and
$\widetilde{\mathbb{B}}$ would  produce one of the degenerate
textures ${\mathbb{C}}_i$ of the original Table~\ref{threeflavor}.
However, even this list is not exhaustive for the degenerate case.
We could have textures which are not  traditionally ordered as
either hierarchial or inverse hierarchial in the three generation
case.  These cases are listed in Table~\ref{threeflavornewalt} and
we call them as alternating textures (see Eq.~(\ref{alttex})).  Thus
in summary, we have covered all possible ways of ordering the three
degenerate eigenvalues in to two sub-matrices, which are not
degenerate themselves.
%--------------------------------
%
%             Table
%
%--------------------------------
\begin{table}[h]
\begin{center}
\begin{tabular}{| l | c  c c  c |}
\hline
&&& &\\
\hspace{0.5cm}{\rm {Mixing $\Rightarrow$}}\hspace{0.5cm} &
\hspace{0.5cm}{\rm {Small}}\hspace{0.5cm} & \hspace{0.5cm}{\rm
{Single maximal}} \hspace{0.5cm} & \hspace{0.5cm}{\rm
{Bimaximal}}\hspace{0.5cm} & \hspace{0.5cm}{\rm
{Tribimaximal}}\hspace{0.5cm}
\\
\hspace{0.5cm}{\rm {$\mathbb{M}_{diag}$}}\hspace{0.5cm} & ${\mathbb{X}}_\ep$ & ${\mathbb{X}}_{SM}$ & ${\mathbb{X}}_{BM}$ & ${\mathbb{X}}_{TBM}$\\
&&&&\\
        \hline
${\mathbb{T}}_1$: Diag[0,1,0] & {\tiny{$\begin{pmatrix}
0 & \ep _{12} & 0 \\
\ep_{12}& 1 & -\ep_{23} \\
\epsilon_{13} & -\ep_{23} & 0
\end{pmatrix}$}}
&
{\tiny{$ \begin{pmatrix} 0 & 0& 0 \\ 0& \frac{1}{2} & -\frac{1}{2}  \\
0 & -\frac{1}{2} & \frac{1}{2} \end{pmatrix} $}}
  &{\tiny{$
  \begin{pmatrix} \frac{1}{2} & \frac{1}{2\sqrt{2}} & -\frac{1}{2\sqrt{2}}
   \\ \frac{1}{2\sqrt{2}} & \frac{1}{4} & -\frac{1}{4}  \\
-\frac{1}{2\sqrt{2}} & -\frac{1}{4} & \frac{1}{4}
\end{pmatrix} $}} & {\tiny{$
\begin{pmatrix}
  \frac{1}{3} & \frac{1}{3} & -\frac{1}{3}
 \\ \frac{1}{3} & \frac{1}{3} & -\frac{1}{3}
  \\-\frac{1}{3} &  -\frac{1}{3}& \frac{1}{3}
  \end{pmatrix} $}}    \\
 ${\mathbb{T}}_2$:
 Diag[1,0,1] & {\tiny{$\begin{pmatrix}
1 & -\ep_{12} & 0 \\ -\ep_{12} & 0 & \epsilon_{23} \\
0 & \epsilon_{23}& 1
\end{pmatrix}$}}
& {\tiny{$ \begin{pmatrix} 1& 0 & 0\\ 0 & \frac{1}{2}& \frac{1}{2}
\\  0 & \frac{1}{2} & \frac{1}{2} \end{pmatrix} $}}  &{\tiny{$
\begin{pmatrix} \frac{1}{2} & -\frac{1}{2\sqrt{2}} &
\frac{1}{2\sqrt{2}}
   \\ -\frac{1}{2\sqrt{2}} & \frac{3}{4} & \frac{1}{4}  \\
\frac{1}{2\sqrt{2}} & \frac{1}{4} & \frac{3}{4}
\end{pmatrix} $}} & {\tiny{$
\begin{pmatrix}
 \frac{2}{3}& -\frac{1}{3} & \frac{1}{3}
 \\ -\frac{1}{3} & \frac{2}{3}& \frac{1}{3}  \\
 \frac{1}{3} & \frac{1}{3}& \frac{2}{3} \end{pmatrix}
  $}}
       \\
\hline
\end{tabular}
\caption[]{\footnotesize{Alternating textures (leading order) for
different mixing scenarios.
 }} \label{threeflavornewalt}
\end{center}
\end{table}
%-----------------------------------------------------------

 \subsection{Case II: Three Sources}
 \label{thsm}
 For more than two seesaw mechanisms at work, the generalisation is straight forward.
 Lets consider the case where there are three sources of neutrino masses. The total
 mass matrix in this case is given by
\begin{equation}
\mathbb{M}_\nu  = \mathbb{M}_\nu^{(1)} + \mathbb{M}_\nu^{(2)} +
\mathbb{M} _\nu^{(3)}~,
\end{equation}
where each of the sub matrices can be thought of having independent
origin through a seesaw mechanism or any other scheme to generate
non-zero neutrino masses.  As with the two-generation case, we will
now consider the case where all the mixings present in each of the
sub matrices are taken to be small and each sub-matrix is assumed to
have only one large eigenvalue. The second assumption is a direct
consequence of assuming that all the three sources contribute
equally and there are no cancellations between the dominant
eigenvalues of the sub-matrices.
 With these assumptions,
the total mass matrix can now be written in terms of the individual
mass matrices as
\begin{equation}
\label{3flsmall} \mathbb{M}_\nu = m_1 \left( \begin{array}{ccc}
\epsilon_{13}^2 & \epsilon_{13} \epsilon_{23} & \epsilon_{13} \\
 \epsilon_{13} \epsilon_{23} & \epsilon_{23}^2 & \epsilon_{23} \\
 \epsilon_{13}  & \epsilon_{23} & 1  \end{array} \right)  +
m_2 \left( \begin{array}{ccc}
\epsilon_{12}^{' 2} & \epsilon'_{12}  &  \epsilon'_{12} \epsilon'_{23}\\
 \epsilon'_{12}  & 1  & -\epsilon'_{23} \\
 \epsilon'_{12} \epsilon'_{23}  &- \epsilon'_{23} & \epsilon_{23}^{' 2}   \end{array} \right)  +
  m_3 \left( \begin{array}{ccc}
1 &- \epsilon''_{12}  & - \epsilon''_{13} \\
 -\epsilon''_{12}  & \epsilon_{12}^{'' 2 } & \epsilon''_{12} \epsilon''_{13} \\
 -\epsilon''_{13}   & \epsilon''_{12} \epsilon''_{13} & \epsilon_{13}^{'' 2}  \end{array} \right)~,
 \end{equation}
where $\epsilon_{ij}, \epsilon'_{ij}, \epsilon''_{ij}$ ($i,j=1,2,3$)
are
 small entries corresponding to small mixing angles in $U^{(i)}_{mix}$.
 This total matrix can be diagonalised by an orthogonal matrix ${\mathbb{O}} \equiv
{\mathbb{O}}_{23} {\mathbb{O}}_{13} {\mathbb{O}}_{12}$, such that
${\mathbb{O}}^T {\mathbb{M}}_{\nu} {\mathbb{O} =
\mbox{Diag}[\mathbb{M}}_{\nu}]$. ${\mathbb{O}}_{ij}$ represents a
rotation in the ${ij}^{th}$ plane.  For example
\begin{equation}
{\mathbb{O}}_{23}  =  \left(
\begin{array}{ccc}
1   & 0  & 0   \\
 0  & \cos \theta_{23}  &\sin \theta_{23}   \\
  0 &- \sin \theta_{23}   &   \cos \theta_{23}
  \end{array}
\right)~.
\end{equation}
\begin{eqnarray}
\theta_{23} &\approx&  {1 \over 2} \tan^{-1}  \left[ { 2 ( m_1
\epsilon_{23} - m_2 \epsilon'_{23} + m_3 \epsilon''_{12}
\epsilon''_{13}) \over m_1 ( 1 - \epsilon_{23}^2 ) -
m_2 ( 1 - \epsilon_{23}^{' 2}) + m_3 (\epsilon_{13}^{'' 2} - \epsilon_{12}^{"2}) }
  \right]~, \nonumber \\
\theta_{13} &\approx&  {1 \over 2} \tan^{-1}  \left[ { 2 (
\widetilde{m}_{13} ) \over \widetilde{m}_{33} -
\widetilde{m}_{11} }    \right]~, \nonumber \\
\theta_{12} &\approx&  {1 \over 2} \tan^{-1}  \left[ { 2 (
\widetilde{m}'_{12} ) \over \widetilde{m}'_{22} -
\widetilde{m}'_{11} }    \right]~,
\end{eqnarray}
where
\begin{eqnarray}
\widetilde{m}_{13}&=& s_{23} (m_1 \ep_{13} \ep_{23} + m_2
\epsilon'_{12} -m_3 \epsilon''_{12})
+ c_{23} (m_1 \ep_{13} +m_2 \epsilon'_{12} \epsilon'_{13} - m_3  \epsilon''_{13})~, \nonumber \\
\widetilde{m}_{33}&=& s_{23} [s_{23} (m_2 +m_1 \epsilon_{23}^2 + m_3
\epsilon''_{12} \epsilon''_{13} )
 +c_{23} (m_1 \ep_{23} -m_2 \epsilon'_{23} +m_3 \epsilon''_{12} \epsilon''_{13}) ] \nonumber \\
&+& c_{23} [ s_{23} (m_1 \epsilon_{23} -m_2 \epsilon'_{23} + m_3
\epsilon''_{12} \epsilon''_{13})
+ c_{23} (m_1+m_2 \epsilon^{' 2}_{23} + m_3 \epsilon^{'' 2}_{13} ) ]~, \nonumber\\
\widetilde{m}_{11}&=&  m_3 + m_1 \epsilon_{13}^2 + m_2 \epsilon_{12}^{' 2}~, \nonumber \\
 \widetilde{m}'_{12}&=&  c_{13} \widetilde{m}_{12} =c_{13} [c_{23} (m_1 \ep_{13} \ep_{23} +
 m_2 \epsilon'_{12} - m_3 \epsilon''_{12}) -s_{23 } (m_1 \ep_{13} + m_2 \epsilon'_{12}
\epsilon'_{13}
 - m_3 \epsilon''_{13})]~, \nonumber \\
\widetilde{m}'_{22}&=& \widetilde{m}_{22} =  c_{23}[ c_{23}(m_2 +m_1
\ep_{23} ^2+ m_3
  \epsilon''_{12} \epsilon''_{13})- s_{23}( m_1 \ep_{23} - m_2 \epsilon'_{23} + m_3 \epsilon''_{12} \epsilon''_{13})]
  \nonumber \\
  &-& s_{23} [c_{23} ( m_1 \ep_{23} - m_2 \epsilon'_{23} + m_3 \epsilon''_{12} \epsilon''_{13})
  -s_{23}(m_1 + m_2 \epsilon_{23} ^{' 2} + m_3 \epsilon_{13}^{'' 2}
  )]~,\nonumber\\
\widetilde{m}'_{11}&=& c_{13} (\widetilde{m}_{11} c_{13} -
\widetilde{m}_{13} s_{13}   ) - s_{13} (\widetilde{m}_{13} c_{13} -
\widetilde{m}_{33} s_{13} ) ~.
\end{eqnarray}
Notice that all the three mass eigenvalues are of the same CP parity
in the above and the degeneracy induced mixing thus works for the
all the three mixing angles. Thus all the three mixing angles are
large.  One can then ask the question whether choosing one of the
mass eigenvalues with a negative CP parity would help in keeping one
of the mixing angles small. The answer is negative, choosing one of
the eigenvalues to have CP parity negative leads to at least two of
the mixing angles to remain small as the degeneracy induced large
mixing mechanism is no longer operative for two of the mixing
angles.  Thus we are back to the case of two seesaw mechanisms which
we have seen in the previous subsection.

While it is possible to visualise GUT models where there are three seesaw
mechanisms at work, it much easier to suitably split a single Type I seesaw mass matrix
 into three sub-matrices. In this case, we can extend Eq.~(\ref{typeIdecomp}) to three
 generations as
\begin{eqnarray}
-{\mathbb M}^{\rm{I}}_\nu &=& m_1 \left( \begin{array}{ccc}
(h_{ee}^D)^2 & h_{ee}^D h_{e \mu}^D & h_{ee}^D h^D_{e\tau}\\
h_{ee}^D h_{e \mu}^D & (h_{e \mu}^D)^2 & h^D_{e\mu} h^D_{e \tau}\\
 h_{ee}^D h^D_{e\tau} & h^D_{e\mu} h^D_{e \tau} & (h^D_{e\tau})^2
  \end{array} \right) +
  %\nonumber \\ &+&
 m_2 \left( \begin{array}{ccc}
(h_{\mu e})^2 & h_{\mu e}^D h_{\mu \mu}^D & h_{\mu e}^D h^D_{\mu \tau}\\
h_{\mu e}^D h_{\mu  \mu}^D & (h_{\mu \mu}^D)^2 & h^D_{\mu \mu} h^D_{\mu \tau}\\
 h_{\mu e}^D h^D_{\mu \tau} & h^D_{\mu \mu} h^D_{\mu \tau} & (h^D_{\mu \tau})^2
  \end{array} \right)  \nonumber\\ &+&
  m_3 \left( \begin{array}{ccc}
(h_{\tau e})^2 & h_{\tau e}^D h_{\tau \mu}^D & h_{\tau e}^D h^D_{\tau \tau}\\
h_{\tau e}^D h_{\tau  \mu}^D & (h_{\tau \mu}^D)^2 & h^D_{\tau \mu} h^D_{\tau \tau}\\
 h_{\tau e}^D h^D_{\tau \tau} & h^D_{\tau \mu} h^D_{\tau \tau} & (h^D_{\tau \tau})^2
  \end{array} \right)~.
  \label{s3}
\end{eqnarray}
Comparing this with Eq.~(\ref{3flsmall}), we see that we will have
three possible solutions for the Yukawa couplings in this case. The
first solution is
\begin{eqnarray}
&& h_{e e}^D ~\sim ~\ep_{13}~, \;\;
 h_{e \mu}^D ~\sim~ \ep_{23}~, \;\;
  h^D_{e \tau} ~\sim ~ \mathcal{O}(1)~, \;\;
  \nonumber\\
  &&
   h_{\mu e}^D ~\sim ~ \ep^\prime_{12}~, \;\;
   h_{\mu \mu}^D ~\sim ~\mathcal{O}(1)~, \;\;
   h_{\mu \tau}^D ~\sim ~ -\ep^\prime_{23}~,\;\;
    \nonumber\\
  &&
     h_{\tau e}^D ~\sim ~ \mathcal{O}(1)~, \;\;
   h_{\tau \mu}^D ~\sim ~ -\ep^{\prime \prime}_{12}~, \;\;
   h_{\tau \tau}^D ~\sim ~ -\ep^{\prime \prime}_{13}~.
\end{eqnarray}
There are two more possibilities given by
\begin{eqnarray}
&& h_{\mu e}^D ~\sim ~\ep_{13}~, \;\;
 h_{\mu \mu}^D ~\sim~ \ep_{23}~, \;\;
  h^D_{\mu \tau} ~\sim ~ \mathcal{O}(1)~, \;\;
  \nonumber\\
  &&
   h_{e e}^D ~\sim ~ \ep^\prime_{12}~, \;\;
   h_{e \mu}^D ~\sim ~\mathcal{O}(1)~, \;\;
   h_{e \tau}^D ~\sim ~ -\ep^\prime_{23}~, \;\;
    \nonumber\\
  &&
     h_{\tau e}^D ~\sim ~ \mathcal{O}(1)~, \;\;
   h_{\tau \mu}^D ~\sim ~ -\ep^{\prime \prime}_{12}~, \;\;
   h_{\tau \tau}^D ~\sim ~ -\ep^{\prime \prime}_{13}~,
\end{eqnarray}
or
\begin{eqnarray}
&& h_{\tau e}^D ~\sim ~\ep_{13}~, \;\;
 h_{\tau \mu}^D ~\sim~ \ep_{23}~, \;\;
  h^D_{\tau \tau} ~\sim ~ \mathcal{O}(1)~, \;\;
  \nonumber\\
  &&
   h_{\mu e}^D ~\sim ~ \ep^\prime_{12}~, \;\;
   h_{\mu \mu}^D ~\sim ~\mathcal{O}(1)~, \;\;
   h_{\mu \tau}^D ~\sim ~ -\ep^\prime_{23}~, \;\;
    \nonumber\\
  &&
     h_{e e}^D ~\sim ~ \mathcal{O}(1)~, \;\;
   h_{e \mu}^D ~\sim ~ -\ep^{\prime \prime}_{12}~, \;\;
   h_{e \tau}^D ~\sim ~ -\ep^{\prime \prime}_{13}~.
\end{eqnarray}
 From the above we see that even if each of the leptonic generation couples
\textit{minimally} with each of the right-handed neutrino,  the
total mixing can be maximal, purely due to the degeneracy
requirement. The comments at the end of subsection~\ref{twsm}
regarding maximally symmetric leptonic mixing matrix hold in this
case too. Finally, note that each set of these solutions is related
by $S_3$ symmetry to the other set.

\section{Summary}
\label{secfour} In the present work, we have concentrated on the
case with two seesaw mechanisms at work which occurs naturally in
many examples like LRS models, SO(10) based GUT models {\textit{etc}}. We
have shown that if both these seesaw mechanisms result in mass
matrices which only have small mixing in them, then the only pattern
of mass eigenvalues which is \textit{naturally} consistent with maximal/large mixing is
the quasi-degenerate pattern for the total mass matrix.

 All the arguments presented in the present work are independent of the details of the sources
 of neutrino masses. However,  depending on the specifics of the model, there could be
 radiative corrections which could significantly modify the mixing angles.  For example,
  if one has  Type I + Type II seesaw mechanism operating at the high scale,
  radiative corrections could significantly modify the mixing angles at the weak scale.
  These effects should be taken in to account when applying the results of the present work to any
particular model.  The impact of radiative corrections, models and
implications for leptogenesis  within this class of
 \textit{hybrid degenerate models} are being studied for a future publication~\cite{ourupcoming}.

\appendix
\section{Generalization of the result for $n$ sources}
\label{appendix} If there are $n$ sources of neutrino masses in a
particular model such that the total mass matrix is given by
\begin{equation}
{\mathbb M}_{\nu} = {\mathbb M}^{{(1)}}_{\nu} + {\mathbb M}^{{(2)}}
_{\nu} + \ldots + {\mathbb M}^{{(n)}} _{\nu}~.
\end{equation}
And further each of the ${\mathbb M}^{{(i)}}_{\nu}$ have  one
dominant diagonal element proportional to its largest  eigenvalue
$m_i$, and rest of the entries to be tiny (all the mixing angles in
all the  ${\mathbb M}^{{(i)}}_{\nu}$ are small); ${\mathbb
M}^{{(i)}}_{\nu}$ are  ordered  in such a way that the ${ii}^{th} $
element is dominant. There are $n$ possible orderings of ${\mathbb
M}^{{(i)}}_{\nu}$. Then the total mass matrix would
\textit{naturally} have a quasi-degenerate pattern with
maximal/large mixing depending  on the number  of \textit{pairs} of
eigenvalues which have the same CP parity, if $m_1 \approx m_2
\approx \ldots  \approx m_n$.  If there are $l$ eigenvalues with the
same CP parity~\footnote{And if the splitting between relevant $m_i$
is smaller than the tiny off-diagonal entries.}, then  $^n
{\cal{C}}_2 + ^{n-l} {\cal{C}}_2 $ (if $(n-l) > 2$) angles will be
large or maximal and the remaining will be small.  An important
exception to the above is the pseudo-Dirac pattern of degenerate
masses, which can only result from a `sum' of two mass matrices both
containing maximal mixing and equal eigenvalues with opposite
ordering in hierarchy. Conversely, at the zeroth order a $n \times
n$ quasi-degenerate matrix with eigenvalues $m_1, m_2, \ldots m_i
\ldots m_n$ (by definition $m_1 \approx m_2 \approx \ldots \approx
m_n$) can be decomposed in to $n$ sub-matrices ${\mathbb
M}^{{(n)}}_{\nu}$, with eigenvalues distributed as
\begin{eqnarray}
{\mathbb M}_{\nu} &=& {\mathbb M}^{{(1)}}_{\nu} + {\mathbb
M}^{{(2)}}
_{\nu} + \ldots + {\mathbb M}^{{(n)}} _{\nu} \nonumber \\
\begin{pmatrix} m_1 & & & & \\
 & m_2  &  && \\
 && \ddots &&\\ &&&& m_n
\end{pmatrix}
&=& \begin{pmatrix} m_1 & & & & \\
 & 0 &  && \\
 && \ddots &&\\ &&&& 0
\end{pmatrix} + \begin{pmatrix} 0 & & & & \\
 & m_2 &  && \\
 && \ddots &&\\ &&&& 0
\end{pmatrix} + \ldots + \begin{pmatrix} 0 & & & & \\
 & 0 &  && \\
 && \ddots &&\\ &&&& m_n
\end{pmatrix}~.
%\nonumber\\
%\{m_1, m_2, \ldots, m_n\}  &= & \{m_1,0,0, \ldots ,0_n\} +
%\{0,m_2,0, \ldots,0_n\} + \ldots + \{0,\ldots  \ldots ,m_n\}~.
 \end{eqnarray}
This holds true irrespective of the mixing present in the  total
mass matrix ${\mathbb M}_{\nu}$.
%-------------------------------------------------------------------------
\acknowledgements{J.C. thanks A. Raychaudhuri for encouragement and
discussions. He also acknowledges support from the Neutrino Project
and RECAPP under the XIth plan of Harish-Chandra Research Institute.
J.C. further acknowledges the hospitality and support from  CHEP, IISc., 
Bangalore where part of the work was carried out. P.M. acknowledges the kind
hospitality received from the Institut f\"{u}r Theoretische Physik
und Astrophysik, Universit\"{a}t  W\"{u}rzburg; Institut f\"{u}r
Theoretische Physik E, RWTH Aachen; CFTP, Instituto Superior
T\'{e}cnico - Universidade T\'{e}cnica de Lisboa as well as the
organisers of ``Workshop towards neutrino technologies" at ICTP,
Italy and ``Lepton Photon 2009" in Hamburg during the final stages
of this work.}
%%%%%%%%%%%%%%%%%%%%%%%
%\bibliographystyle{apsrev}

%%%%%%%%%%%%%%%%%%%%%%%%

\end{document}